\documentclass[10pt,english,prb,nofootinbib]{revtex4-2}

\usepackage[T1]{fontenc}
\usepackage[latin9]{inputenc}
\usepackage{color}
\usepackage{bm}
\usepackage{amsmath}
\usepackage{graphicx}

\makeatletter

\providecommand{\tabularnewline}{\\}

\usepackage{babel}

\makeatother

\usepackage{babel}
\begin{document}
\title{Low-frequency signature of magnetization nutation in nanomagnets}
\author{M. P. Adams$^{1}$}
\email[Electronic address:]{ michael.adams@uni.lu}

\author{R. Bastardis$^{2}$}
\email[Electronic address:]{ roland.bastardis@univ-perp.fr}

\author{A. Michels$^{1}$}
\email[Electronic address:]{ andreas.michels@uni.lu}

\author{H. Kachkachi$^{2}$}
\email[Electronic address:]{ hamid.kachkachi@univ-perp.fr}

\affiliation{$^{1}$Department of Physics and Materials Science, University of
Luxembourg, 162A~avenue de la Faiencerie, L-1511 Luxembourg, Grand
Duchy of Luxembourg}
\affiliation{$^{2}$Universit\'e de Perpignan via Domitia, Lab. PROMES CNRS UPR8521,
Rambla de la Thermodynamique, Tecnosud, 66100 Perpignan, France}
\begin{abstract}
In this work, we show that surface anisotropy in nanomagnets induces
a nutational motion of their magnetization at various frequencies,
the lowest of which can be described by the macrospin model whose
dynamics is governed by an effective energy potential. We derive analytical
expressions for the precession and nutation frequencies and amplitudes
as functions of the size of the nanomagnet and its atomistic parameters,
such as the exchange coupling and the onsite anisotropy. Our analytical
model predicts a reduction of the precession frequency with increased
surface anisotropy. We also simulate the dynamics of the corresponding
atomistic many-spin system and compare the results with the effective
model. We thereby show that the first nutation mode induced by the
finite size and surface anisotropy occurs at a frequency that is four
times larger than the precession frequency, thus lending itself to
a relatively easy detection by standard experiments of magnetic resonance. 
\end{abstract}
\maketitle

\section{introduction}

Nutation is the well known motion of a gyroscope described by classical
mechanics; it develops whenever an external force tilts the rotation
axis away from the direction of the gravity field, since then the
rotation axis no longer coincides with the direction of the angular
momentum.

In the case of a magnetic material, magnetization nutation occurs
whenever a time-dependent magnetic field (rf or microwave field) is
present in the system. Accordingly, it has been demonstrated that
there appears the fundamental effect of transient nutations in NMR~\citep{Torrey49pr},
EPR~\citep{verfes73jcp,atkinsEtal74cpl,fedoruk02jas} and optical
resonance~\citep{hoctan68prl}. Spin nutation was first predicted
in Josephson junctions~\citep{zhufra06jpcm,franzhu08njp,franson08nanotech,PhysRevB.71.214520,PhysRevLett.92.107001}
and was later developed using various approaches based on relativistic
quantum mechanics, first principles~\citep{mondaletal17prb,monopp_jpcm20},
and electronic structure theory~\citep{PhysRevLett.108.057204,PhysRevB.84.172403,PhysRevB.92.184410,thoningetal17sp,chengetal17prb}.
A more recent macroscopic approach~\citep{ciorneietal11prb,oliveetal12apl,doi:10.1063/1.4921908,oliweg16jpcm}
deals with the magnetization nutation by accounting for magnetic inertia
through the addition of a second-order time derivative of the magnetization
in the Landau-Lifshitz-Gilbert equation~\citep{lanlif35,gilbert56phd}.
On the experimental side, evidence of the effects of nutation has
been reported for thin films~\citep{LiEtAl_prb15,NeerajEtAl_nat21,Unikandanunni_prl22, DeEtAl_arxiv24}
in the THz regime. Indeed, magnetization nutation meets with a strong
and expanding interest within the magnetism community owing to the
real possibility of addressing and controlling the ultra-fast magnetization
dynamics, in view of potential practical applications in high-speed
data processing and storage technologies based on magnetic materials.
Considering the strong potential of achieving high storage densities,
this issue acquires still more enthusiasm for nanoscaled magnetic
systems.

The dynamic behavior of nanomagnets requires a good understanding
of the intricate interplay between different factors influencing the
spin dynamics, such as finite-size, boundary, and surface effects.
In particular, surface anisotropy emerges as a crucial parameter that
can significantly influence the dynamics of nanomagnets as it affects
the potential energy and thereby the relaxation rates~\citep{dejardinetal08jpd,Vernay_etal_acsucept_PRB2014}.
Recently, it was shown~\citep{bothen_prb12,basvenkac_prb18} that
spin misalignment, induced by surface effects, triggers nutational
motion of the magnetization of a nanomagnet, with frequencies ranging
from tens of GHz to THz, depending on various physical parameters,
the predominant among them being the spin-spin exchange coupling and
the on-site surface anisotropy. For example, in Refs.~\citep{BastardisEtal-jpcm2017,basvenkac_prb18}
it was shown that surface anisotropy in nanocubes triggers absorption
modes of higher frequency than the ordinary uniform precessional motion
with frequencies $f_{\mathrm{p}}$$\sim$$10\,\mathrm{GHz}$ (conventional
ferromagnetic resonance). More precisely, the authors of Ref.~\citep{basvenkac_prb18}
observed a resonance at a frequency $f_{\mathrm{c}}$ that is 4 times
$f_{\mathrm{p}}$ and another frequency $f_{\mathrm{n}}$ in the THz
regime. While $f_{\mathrm{n}}$ is related to atomic spin fluctuations
and their exchange coupling, the frequency $f_{\mathrm{c}}$ emerges
from the cubic anisotropy induced by spin inhomogeneities at the surface
of the nanomagnet~\citep{garkac03prl,kacbon06prb,yanesetal07prb}.
As such, $f_{\mathrm{c}}$ corresponds to the nutational mode with
the lowest frequency and should be more easily detected in resonance
experiments on, \emph{e.g.} arrays of nanomagnets with acute surface
effects, such as platelets, pillars, spheres, or cubes.

In the present work, we focus on the frequency $f_{\mathrm{c}}$ and
derive its analytical expression using the effective macrospin model
whose potential energy is a polynomial in the components of the net
magnetic moment, with coefficients that are functions of the spin-spin
exchange coupling and on-site anisotropy constants~\citep{garkac03prl,kacbon06prb,yanesetal07prb,kachkachi07j3m}.
The results obtained using the effective model are compared with those
of numerical simulations for the corresponding many-spin nanomagnet.
The effective model, thus validated, allows us to derive analytical
expressions for the precession and nutation frequencies and the nutation
amplitude and to study them in terms of the nanomagnet size and surface
anisotropy.

Plan of the paper:~We first define the many-spin and macrospin approaches
and establish the connection between them. Then, we discuss the numerical
results from both models for the time trajectories of the net magnetic
moment. In particular, we discuss the results regarding the magnetization
nutation and its dependence on the particle size and surface anisotropy.
This study allows us to validate the effective model which is then
used to derive the analytical expressions for the precession and nutation
frequencies and amplitudes. From these expressions and further simulations,
we draw the main conclusions of the present work.

\section{Model for the magnetization dynamics}

\subsection{Many-spin approach (MSP)}

In the many-spin problem (MSP) or many-spin approach, or still atomistic
approach, the nanomagnet is regarded as a crystallite of a given shape
and size, containing $\mathcal{N}$ atomic spins $\boldsymbol{S}_{i}=\mu_{\mathrm{a}}\boldsymbol{s}_{i}$
located at the nodes of a given crystalline structure; $\mu_{\mathrm{a}}$
denotes the atomic magnetic moment and $\boldsymbol{s}_{i}$ is a
unit vector in the direction of $\boldsymbol{S}_{i}$. The magnetic
state of such a system is studied with the help of the atomistic Hamiltonian~\citep{dimwys94prb,kodber99prb,kacgar01physa300,kacgar01epjb,igllab01prb,kacdim02prb,kacgar05springer,kazantsevaetal08prb,AdamsEtal_jac22_ana,AdamsEtAl_prb24}:
\begin{align}
\mathcal{H} & =\mathcal{H_{\mathrm{exch}}}+\mathcal{H}_{\mathrm{\mathrm{Z}}}+\mathcal{H}_{\mathrm{A}}\nonumber \\
 & =-\frac{1}{2}J\sum_{\left\langle i,j\right\rangle }\boldsymbol{s}_{i}\cdot\boldsymbol{s}_{j}-\mu_{\mathrm{a}}\mathbf{B}_{0}\cdot\sum_{i=1}^{\mathcal{N}}\boldsymbol{s}_{i}+\sum_{i=1}^{\mathcal{N}}\mathcal{H}_{\mathrm{A},i},\label{eq:Ham-MSP}
\end{align}
where $\mathcal{H_{\mathrm{exch}}}$ is the nearest-neighbor (n.n.)
exchange energy with $J>0$, $\mathcal{H}_{\mathrm{\mathrm{Z}}}$
denotes the Zeeman energy with $\mathbf{B}_{0}=\mu_{0}\mathbf{H}_{0}$
being the homogeneous externally applied magnetic field, and $\mathcal{H}_{\mathrm{A}}$
the magnetic anisotropy energy. For the core spins, we assume the
anisotropy to be of uniaxial symmetry with constant $K_{\mathrm{c}}$,
while for surface spins we adopt the model proposed by N\'eel~\citep{nee54jpr}
with constant $K_{\mathrm{s}}$. $\mathcal{H}_{\mathrm{A},i}$ is
then given by 
\begin{equation}
\mathcal{H}_{\mathrm{A},i}=\begin{cases}
-K_{\mathrm{c}}\left(\boldsymbol{s}_{i}\cdot\mathbf{e}_{z}\right)^{2}, & i\in{\mathrm{core}}\\
+\frac{1}{2}K_{\mathrm{s}}{\sum_{j\in{\mathrm{n.n.}}}}\left(\boldsymbol{s}_{i}\cdot\mathbf{u}_{ij}\right)^{2}, & i\in\mathrm{surface},
\end{cases}\label{eq:HamUA-NSA}
\end{equation}
where the unit vectors $\mathbf{u}_{ij}=(\mathbf{r}_{i}-\mathbf{r}_{j})/\|\mathbf{r}_{i}-\mathbf{r}_{j}\|$
connect the nearest-neighbor spins ``$i$'' and ``$j$''. All
physical parameters are measured in units of energy per atom.

%Here, $\mathrm{n.n.}$ stands for nearest neighbors. Not needed (see above).

The (undamped) dynamics of the many-spin system is governed by the
following Larmor equation: 
\begin{equation}
\frac{d\boldsymbol{s}_{i}}{d\tau}=\boldsymbol{s}_{i}\times\bm{b}_{{\rm eff},i}\label{eq:LarmoEqt}
\end{equation}
with the (normalized) local effective field $\bm{b}_{{\rm eff},i}$
acting on $\bm{s}_{i}$ being defined by $\bm{b}_{{\rm eff},i}=-\delta\mathcal{H}/\delta\bm{s}_{i}$;
$\tau$ is the reduced time defined by: 
\begin{equation}
\tau\equiv\frac{t}{\tau_{{\rm s}}},\label{eq:ReducedTime}
\end{equation}
where $\tau_{{\rm s}}=\mu_{\mathrm{a}}/\left(\gamma J\right)$ is
a characteristic time of the system's dynamics. By way of example,
for cobalt $J=8{\rm \ meV}$, leading to $\tau_{{\rm s}}=70\,{\rm fs}$.
Henceforth, we will only use the dimensionless time $\tau$. Accordingly,
the frequency $\omega=2\pi f=2\pi/T$ is measured in units of $\tau_{{\rm s}}^{-1}$
and, as such, in the sequel $\nu\equiv\tau_{{\rm s}}\omega$. Note
that in these units, $b_{{\rm eff}}$ is equal to the effective field
(in Tesla) multiplied by $\mu_{\mathrm{a}}/J$ and is thus dimensionless.

To study the dynamics of this MSP, for arbitrary values of all physical
parameters, one resorts to numerical methods for solving Eq.~(\ref{eq:LarmoEqt})
using, for instance, the iterative routine based on the $4^{\mathrm{th}}$-order
Runge-Kutta scheme in combination with the projection step $\boldsymbol{s}_{i}^{\nu+1}=\boldsymbol{s}_{i}^{\mathrm{RK4}}/\|\boldsymbol{s}_{i}^{\mathrm{RK4}}\|$
to preserve the constraint $\|\boldsymbol{s}_{i}\|=1$. The net magnetic
moment $\boldsymbol{m}$ (unit vector) is then computed using 
\begin{equation}
\boldsymbol{m}=\frac{\sum_{i}\boldsymbol{s}_{i}}{\|\sum_{i}\boldsymbol{s}_{i}\|}.\label{eq:NetMagMoment}
\end{equation}
For the initial state of the system, we choose a coherent spin configuration
along the orientation $\left(\theta_{0},\phi_{0}\right)$, \emph{i.e.}
$\bm{s}_{i}=\left(\sin\theta_{0}\cos\phi_{0},\sin\theta_{0}\sin\phi_{0},\cos\theta_{0}\right)$,
for all $i=1,\ldots,\mathcal{N}$.

Nutational motion was investigated in Ref.~\citep{basvenkac_prb18}
using this MSP model and the results were compared with those of the
macrospin model studied with the help of the augmented Landau-Lifshitz-Gilbert
equation. In the present work, our aim is to derive analytical expressions
for the precession and nutation frequencies in terms of the atomistic
parameters $J,K_{\mathrm{c}},K_{\mathrm{s}}$ and the size of the
nanomagnet. For this purpose, we resort to the effective one-spin
problem derived in Refs.~\citep{garkac03prl,kacbon06prb,yanesetal07prb,kachkachi07j3m}
and compare the corresponding results with those obtained for the
MSP in Ref.~\citep{basvenkac_prb18} in regard with the nutational
motion.

%The exchange coupling and anisotropy constants are measured in Joule per atom. Not needed (see above).

\subsection{Effective macrospin approach (EOSP)}

The effective macrospin approach, or effective one-spin problem (EOSP),
is obtained from the many-spin problem under certain conditions regarding
the surface anisotropy, which should not be too strong with respect
to the exchange coupling, and the particle size, which should not
be too small (see Refs.~\citep{garkac03prl,kacbon06prb,kachkachi07j3m,yanesetal07prb}
for details). The EOSP consists of the net magnetic moment Eq.~(\ref{eq:NetMagMoment})
evolving in an effective energy potential given by 
\begin{equation}
\mathcal{H}_{\mathrm{eff}}\simeq-k_{2}m_{z}^{2}+k_{4}\sum_{\alpha=x,y,z}m_{\alpha}^{4}.\label{eq:EffPotential}
\end{equation}
The values and signs of the (effective) coefficients $k_{2}$ and
$k_{4}$ are functions of the atomistic parameters $J,K_{\mathrm{c}},K_{\mathrm{s}}$,
in addition to the size and shape of the nanomagnet and its underlying
crystal lattice~\citep{yanesetal07prb}. In the following, we will
use the dimensionless constants defined by $k_{\mathrm{c}}=K_{\mathrm{c}}/J$
and $k_{\mathrm{s}}=K_{\mathrm{s}}/J$, so that $k_{2}$ and $k_{4}$
are dimensionless.

For a nanomagnet cut out of a simple cubic lattice, we have~\citep{garkac03prl,garanin_prb18}
\begin{equation}
k_{2}=k_{\mathrm{c}}\frac{N_{\mathrm{c}}}{\mathcal{N}},\quad k_{4}\simeq\begin{cases}
\kappa\frac{k_{\mathrm{s}}^{2}}{z}, & \mathrm{sphere},\\
\\
\left(1-0.7/\mathcal{N}^{1/3}\right)^{4}\frac{k_{\mathrm{s}}^{2}}{z}, & \mathrm{cube},
\end{cases}\label{eq:k2k4}
\end{equation}
where $z=6$ is the coordination number, $\kappa$ represents a (dimensionless)
surface integral, and $N_{\mathrm{c}}$ is the number of atoms in
the core of the nanomagnet (with full coordination), while $\mathcal{N}$
is the total number of atoms (including both the core and surface).
Note that $\kappa$ was derived in Ref.~\onlinecite{garkac03prl}
in the absence of core anisotropy. As shown in Ref. \citep{AdamsEtal_ps23},
in the presence of the latter, the spin misalignment caused by surface
anisotropy does not propagate to the center of the nanomagnet; it
is ``fended off'' by the uniaxial anisotropy in the core which tends
to align all spins together. The result of this competition is that
$k_{4}$ (or $\kappa$) is multiplied by the factor $N_{\mathrm{c}}/\mathcal{N}$.
Likewise, for both the cube and sphere, $k_{2}$ may be approximated~\citep{kacbon06prb}
by the first expression in Eq.~(\ref{eq:k2k4}).

From the effective Hamiltonian in Eq.~(\ref{eq:EffPotential}), we
derive the following effective field that drives the dynamics of the
magnetic moment $\boldsymbol{m}$
\begin{equation}
\boldsymbol{b}_{\mathrm{eff}}=2k_{2}m_{z}\boldsymbol{e}_{z}-4k_{4}\left(m_{x}^{3}\boldsymbol{e}_{x}+m_{y}^{3}\boldsymbol{e}_{y}+m_{z}^{3}\boldsymbol{e}_{z}\right)\label{eq:EOSP-EffectiveField}
\end{equation}
through the Larmor equation of motion for $\boldsymbol{m}$ as follows:
\begin{equation}
\frac{d\boldsymbol{m}}{d\tau}=\boldsymbol{m}\times\boldsymbol{b}_{\mathrm{eff}}.\label{eq:NetMagMom_LarmorEqt}
\end{equation}

We would like to emphasize that \textcolor{black}{the EOSP described
here is not some variant of a single-domain macrospin model. }Indeed,
it is a macrospin model, but with coefficients that are functions
of the atomistic physical parameters. In other terms, it has the advantage
to be a \textquotedbl simple\textquotedbl{} macroscopic model which
inherits (to some extent) the nanomagnet atomistic features, especially
spin inhomogeneities.

\section{magnetization nutation}

\subsection{MSP versus EOSP}

\begin{figure*}
\begin{centering}
\includegraphics[scale=0.3]{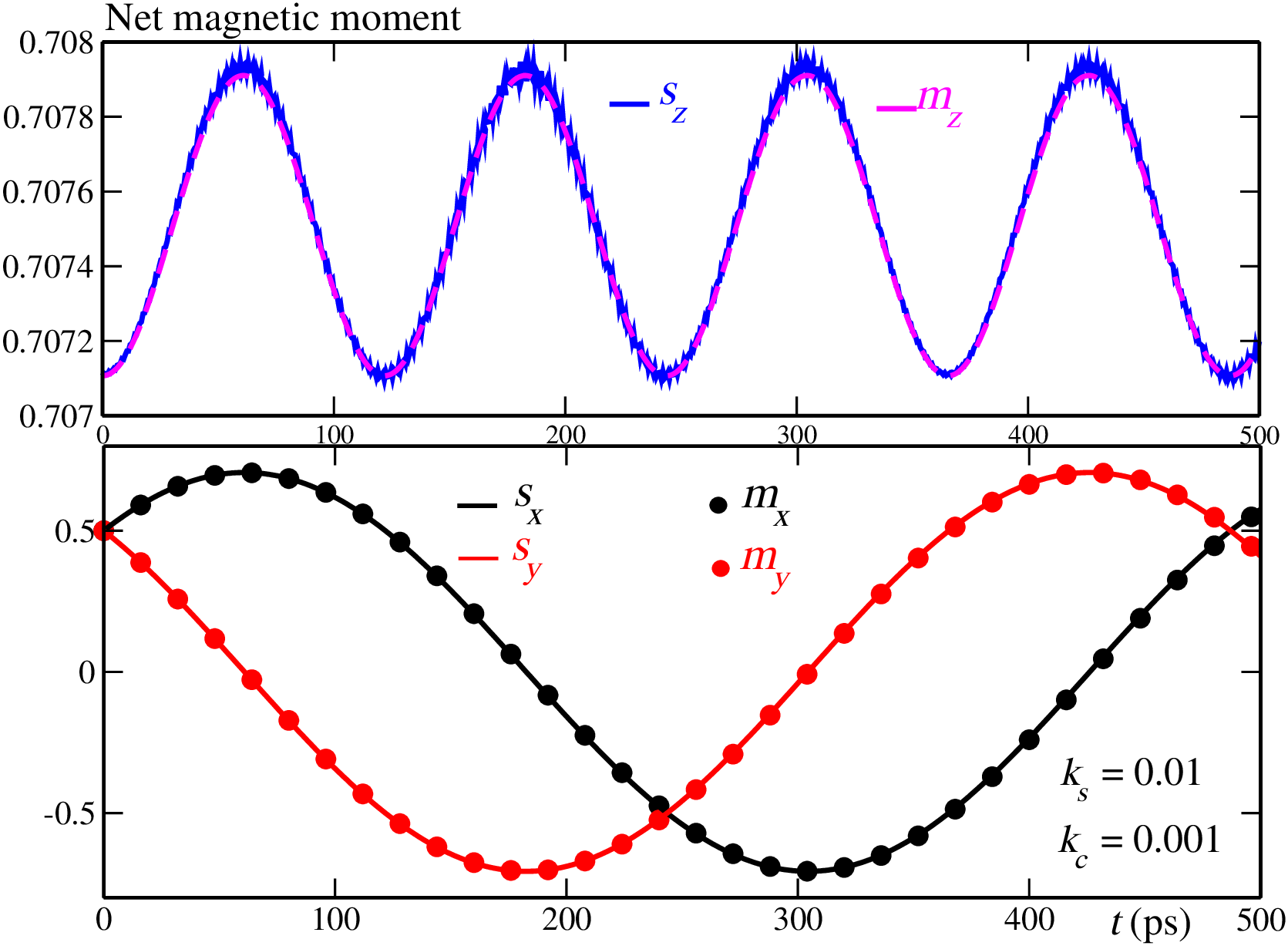} \includegraphics[scale=0.3]{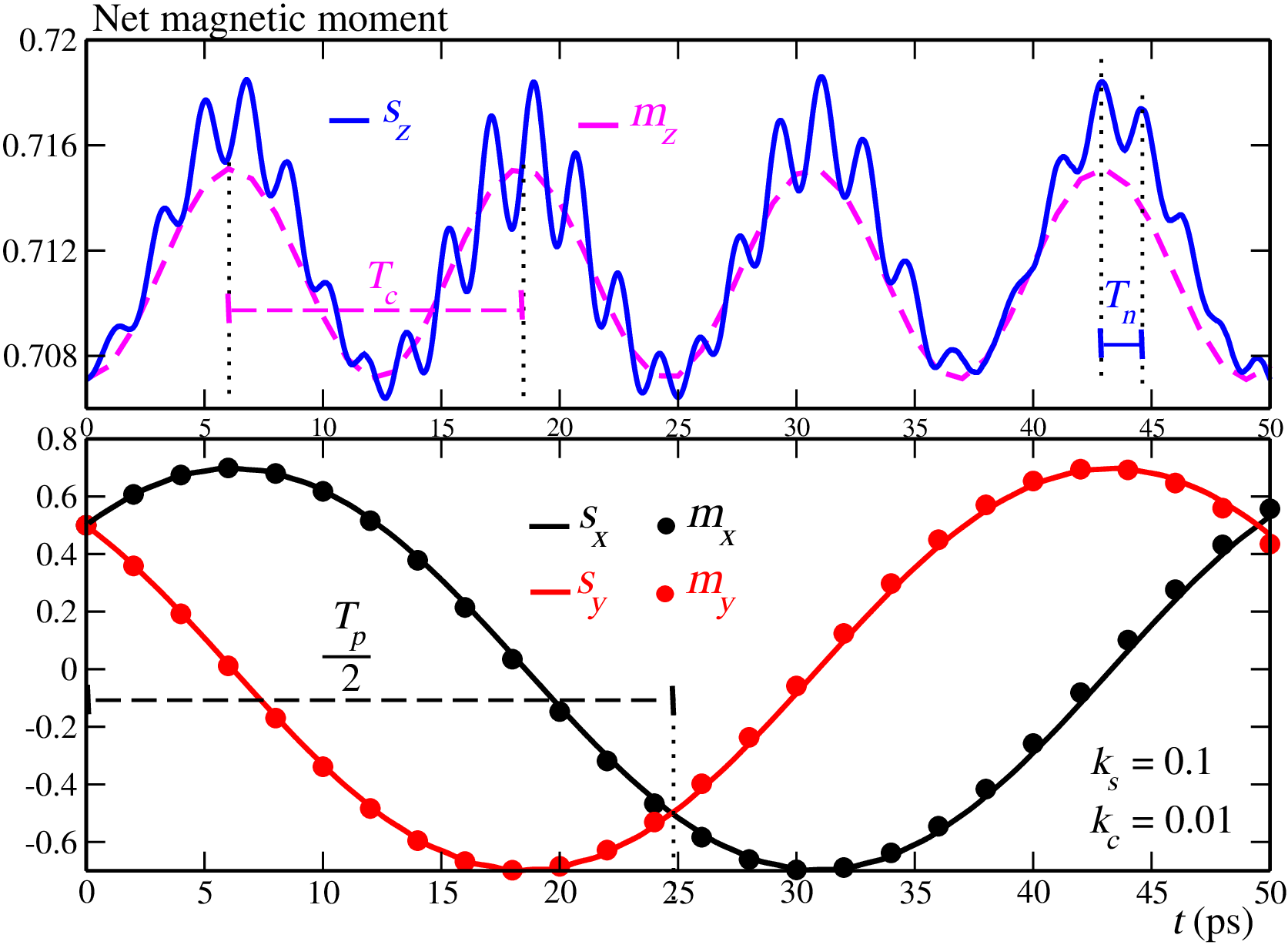}
\par\end{centering}
\caption{\label{fig:Spheres}Components of the net magnetic moment for both
MSP (continued lines) and EOSP (dashed lines and symbols) for the spherical
nanomagnet with a sc lattice and size $\mathcal{N}=1088$. The constants
$k_{\mathrm{c}}$ and $k_{\mathrm{s}}$ are indicated in the legend
and the effective coefficients are given in Table~\ref{tab:spheres}.
On the right, $T_{\mathrm{p}},T_{\mathrm{c}},T_{\mathrm{n}}$ denote
the periods corresponding to the respective frequencies $f_{\mathrm{p}},f_{\mathrm{c}},f_{\mathrm{n}}$
mentioned in the text.}
\end{figure*}

\begin{table}[h]
\begin{centering}
\begin{tabular}{|c|c|c|c|}
\hline 
$\mathcal{N}=1088$ ($N_{\mathrm{c}}=706)$  & $k_{2}$  & $k_{4}$  & $\omega_{\mathrm{c}}$(GHz)\tabularnewline
\hline 
\hline 
$k_{\mathrm{c}}=0.001,k_{\mathrm{s}}=0.01$  & $6.5\times10^{-4}$  & $5.8\times10^{-6}$  & 51.5\tabularnewline
\hline 
$k_{\mathrm{c}}=0.01,k_{\mathrm{s}}=0.1$  & $6.5\times10^{-3}$  & $5.81\times10^{-4}$  & 515\tabularnewline
\hline 
\end{tabular}
\par\end{centering}
\caption{\label{tab:spheres}Effective parameters used for the simulations
on the nanospheres {[}from Eq.~(\ref{eq:k2k4}){]}.}
\end{table}

\begin{figure*}
\begin{centering}
\includegraphics[scale=0.3]{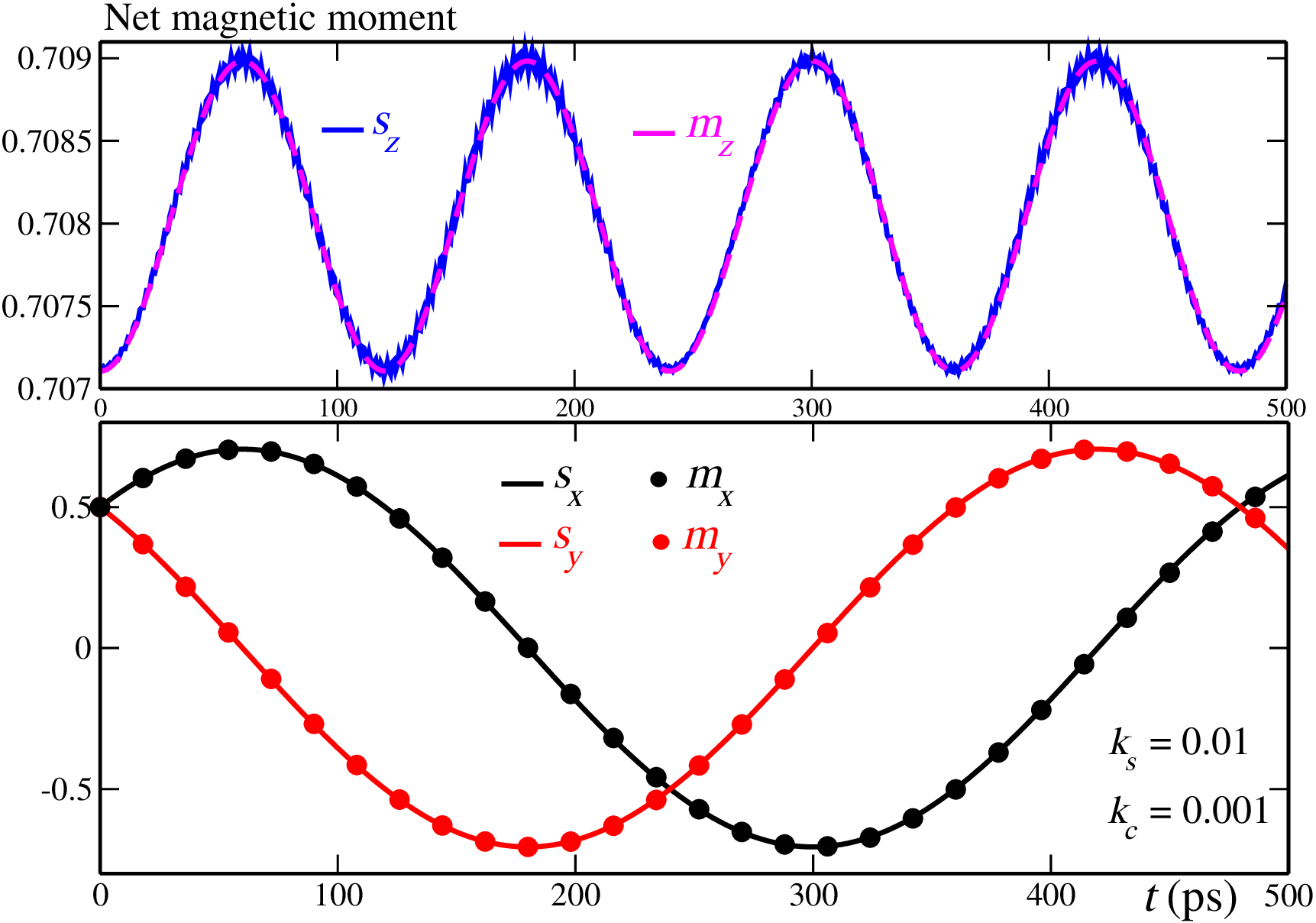}\includegraphics[scale=0.3]{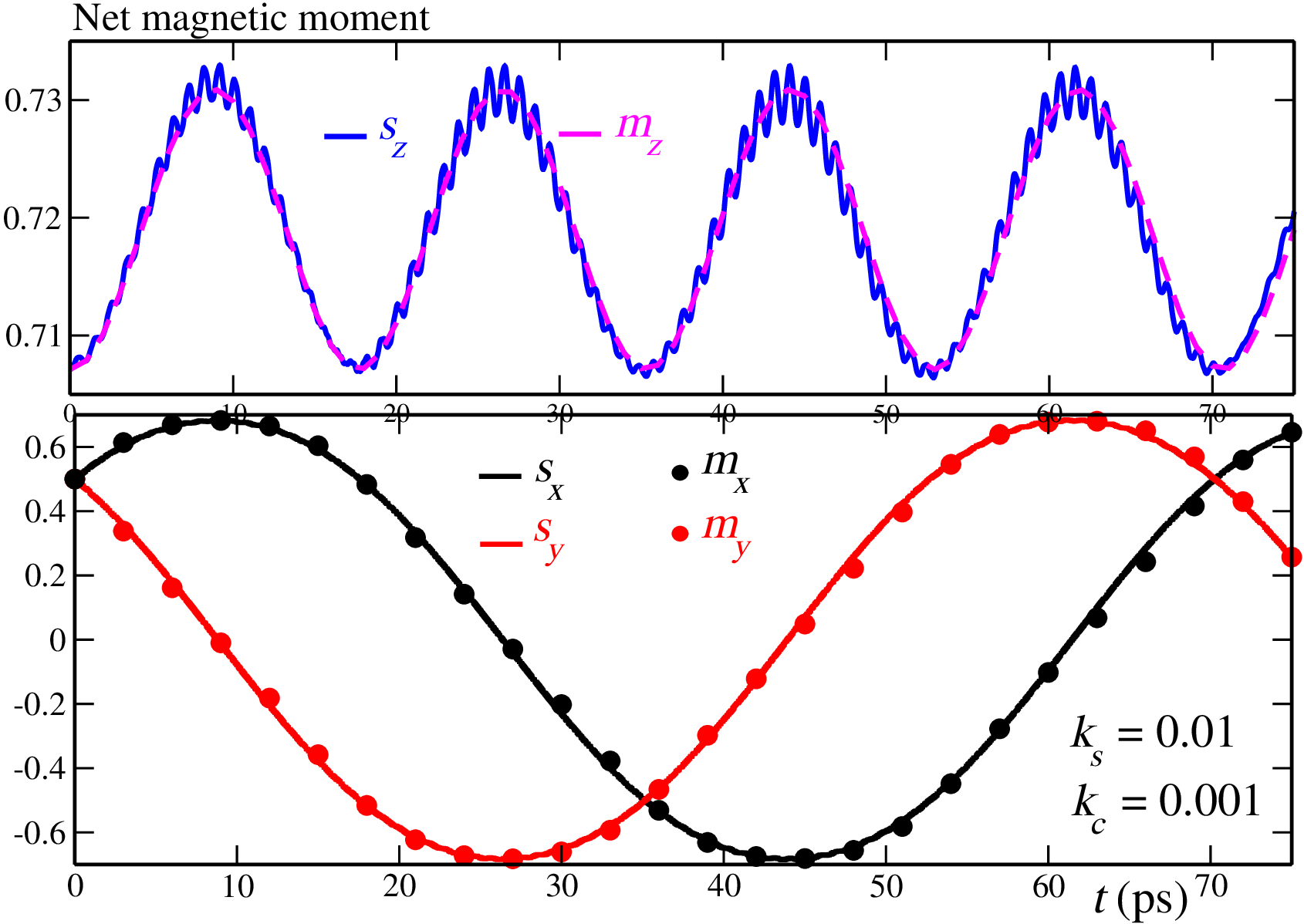}
\par\end{centering}
\caption{\label{fig:Cubes}Components of the net magnetic moment for both MSP
(continued lines) and EOSP (dashed lines and symbols) for a cube-shaped
nanomagnet with a sc lattice and sizes $\mathcal{N}=15^{3}$ (left)
and $\mathcal{N}=9^{3}$ (right). The constants $k_{\mathrm{c}}$
and $k_{\mathrm{s}}$ are indicated in the legend and the effective
coefficients are given in Table~\ref{tab:nanocubes}. }
\end{figure*}

\begin{table}[h]
\begin{centering}
\begin{tabular}{|c|c|c|c|c|}
\hline 
$\mathcal{N}$  & $N_{\mathrm{c}}$  & $k_{2}$  & $k_{4}$  & $\omega_{\mathrm{c}}$(GHz)\tabularnewline
\hline 
\hline 
$729=9^{3}$  & 343  & $4.7\times10^{-4}$  & $12\times10^{-6}$  & 355\tabularnewline
\hline 
$3375=15^{3}$  & 2197  & $6.5\times10^{-4}$  & $13.8\times10^{-6}$  & 52\tabularnewline
\hline 
\end{tabular}
\par\end{centering}
\caption{\label{tab:nanocubes}Effective parameters used for the simulations
on the nanocubes {[}from Eq.~(\ref{eq:k2k4}){]}.}
\end{table}

\begin{figure*}
\begin{centering}
\includegraphics[scale=0.3]{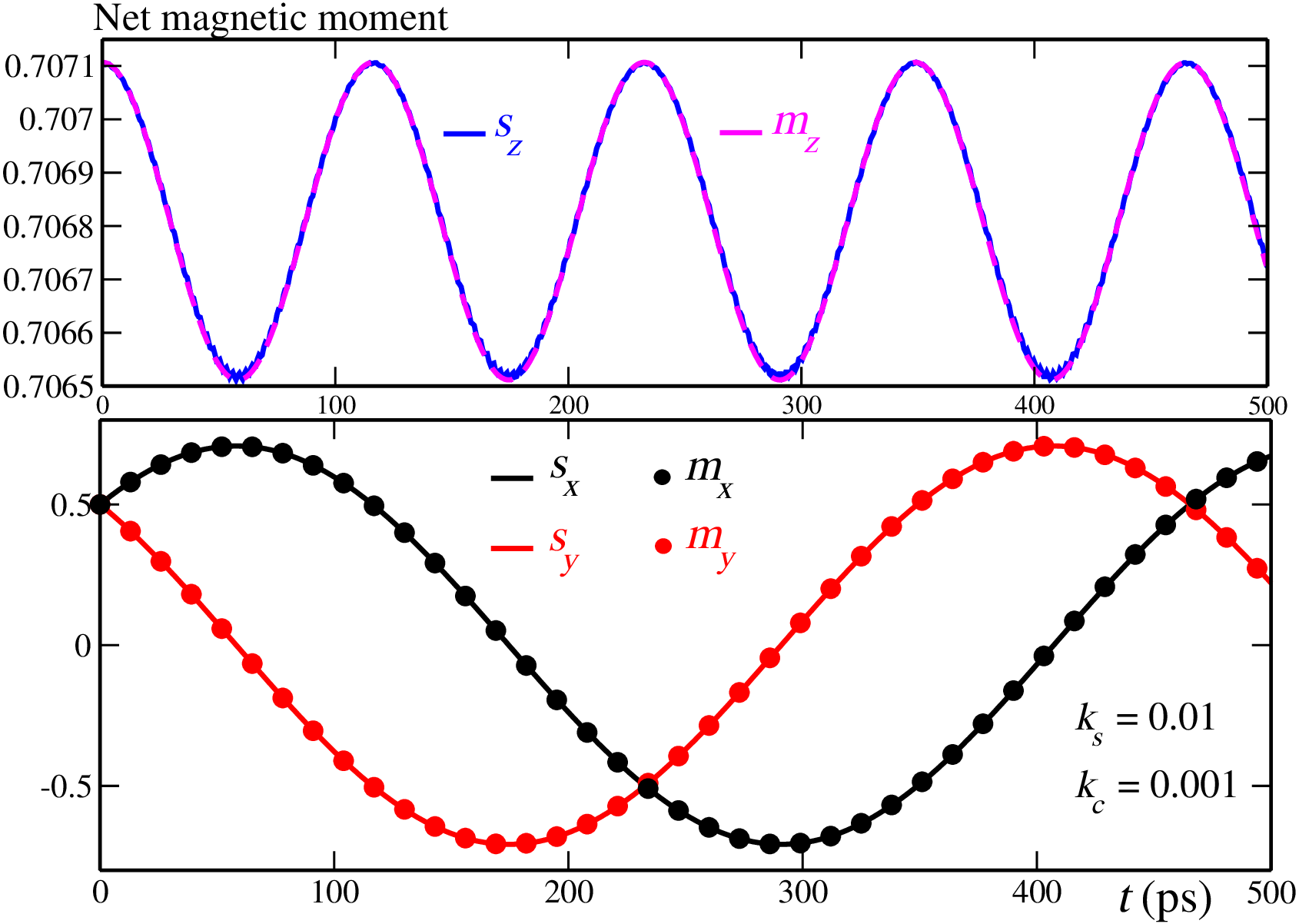}\includegraphics[scale=0.3]{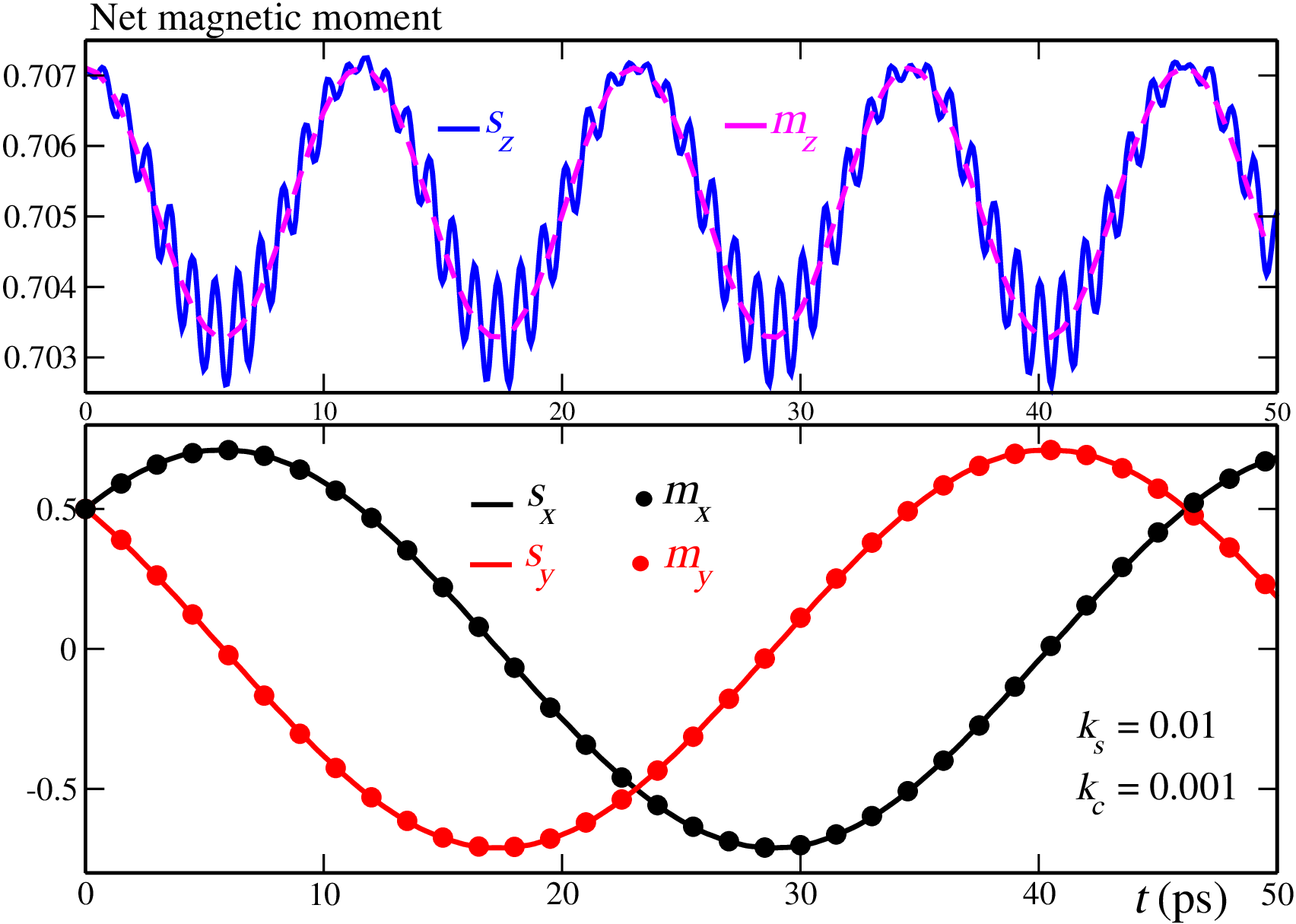}
\par\end{centering}
\caption{\label{fig:Cubo_octahedron}Components of the net magnetic moment
for both MSP (continued lines) and EOSP (dashed lines and symbols)
for the truncated octahedron with fcc lattice and size $\mathcal{N}=2075$.
The constants $k_{\mathrm{c}}$ and $k_{\mathrm{s}}$ are indicated
in the legend and the effective coefficients are given in Table~\ref{tab:CuboOctah}.}
\end{figure*}

\begin{table}[h]
\begin{centering}
\begin{tabular}{|c|c|c|c|}
\hline 
$\mathcal{N}=2075$  & $k_{2}$  & $k_{4}$  & $\omega_{\mathrm{c}}$(GHz)\tabularnewline
\hline 
\hline 
$k_{\mathrm{c}}=0.001,k_{\mathrm{s}}=0.01$  & $6.7\times10^{-4}$  & $-4.5\times10^{-6}$  & 54.6\tabularnewline
\hline 
$k_{\mathrm{c}}=0.01,k_{\mathrm{s}}=0.1$  & $6.7\times10^{-3}$  & $-2.9\times10^{-4}$  & 546\tabularnewline
\hline 
\end{tabular}
\par\end{centering}
\caption{\label{tab:CuboOctah}Effective parameters used for the simulations
on the truncated octahedrons obtained by a fitting procedure. Note
that $k_{4}$ is negative for the fcc lattice~\citep{yanesetal07prb}.}
\end{table}

In this section, we compare the dynamics of the net magnetic moment
of different many-spin nanomagnets with that of the corresponding
effective models for spheres~\citep{SalzemanEtAl_csa23}, cubes~\citep{snoecketal08nl,trunovaetal08jap,jiangetal10jac,mehdaouietal10j3m,kronastetal11nl,okellyetal12nanotech,BastardisEtal-jpcm2017}
and truncated octahedrons~\citep{jametetal01prl,jametetal04prb,yanesetal07prb}.
Before discussing the results in more detail, let us summarize the
procedure followed for the comparison between the MSP and EOSP. For
a given nanomagnet with a given size, shape, underlying lattice, and
energy parameters ($J,K_{\mathrm{c}},K_{\mathrm{s}}$), we solve the
undamped Larmor equation (\ref{eq:LarmoEqt}) for all atomic spins
$\bm{s}_{i}$ starting from the initial state $\left(\theta_{0},\phi_{0}\right)$
taken here to be $\left(\pi/4,\pi/4\right)$. Then, we plot the components
$m_{\alpha}$ (with $\alpha=x,y,z$) of the net magnetic moment (of
the MSP) as functions of time\footnote{Note that in all plots we denote by $s_{\alpha}$, $\alpha=x,y,z$,
the Cartesian components of the net magnetic moment of the MSP and
by $m_{\alpha}$, $\alpha=x,y,z$, those of the magnetic moment of
the EOSP.}. Next, for a cube or a sphere with a simple cubic lattice, we use
the expressions for $k_{2}$ and $k_{4}$ in Eqs.~(\ref{eq:k2k4})
to solve the Larmor Eq.~(\ref{eq:NetMagMom_LarmorEqt}) for the EOSP
and obtain the components $m_{\alpha}$ as functions of time. For
other shapes and/or lattice structures, such as the truncated octahedron
or cubes with \emph{e.g. }body-centered-cubic lattice {[}see Fig.
\ref{fig:Cubo_octahedron}{]}, $k_{4}$ is obtained by fitting the
MSP data. Note that in the case of an underlying bcc or fcc lattice,
$k_{4}$ becomes negative~\citep{yanesetal07prb}.

The results of such a comparison confirm the following facts :~the
net magnetic moment of the many-spin nanomagnet exhibits three oscillation
modes with different frequencies {[}see Fig. \ref{fig:Spheres} right{]}:
the precession frequency $f_{\mathrm{p}}$ for the $x,y$ components
and the nutation frequencies $f_{\mathrm{c}}$ and $f_{\mathrm{n}}$
for the $z$ component. The data in Figs.~\ref{fig:Spheres}, \ref{fig:Cubes},
\ref{fig:Cubo_octahedron} are plotted as continued lines, in black
and red for $s_{x,y}$ and in blue for $s_{z}$. The corresponding
EOSP magnetic moment exhibits only two oscillation frequencies:~the
precession frequency $f_{\mathrm{p}}$ for $m_{x,y}$ (in red and
black symbols) and the (smaller) nutation frequency $f_{\mathrm{c}}$
for $m_{z}$ (in magenta). This result obtains in all cases of size,
shape and other physical parameters, as confirmed in Figs.~\ref{fig:Spheres},
\ref{fig:Cubes} and \ref{fig:Cubo_octahedron}. We see that as far
as $f_{\mathrm{p}}$ and $f_{\mathrm{c}}$ are concerned, and this
corresponds to the validity domain of the EOSP (i.e.\emph{, }relatively
weak surface anisotropy), the dynamics of EOSP perfectly matches that
of the MSP. However, the latter exhibits (in blue) an extra wiggling
motion in the component $s_{z}$ (with frequency $f_{\mathrm{n}}$)
on top of a signal with frequency $f_{\mathrm{c}}$. As mentioned
in the introduction and discussed in Ref.~\citep{basvenkac_prb18},
the appearance of the frequency $f_{\mathrm{c}}$ is due to the quartic
term in the Hamiltonian (\ref{eq:EffPotential}), which is a consequence
of spin disorder induced by surface anisotropy in the EOSP regime;
the coefficient of this term ($k_{4}$) is a function of $K_{\mathrm{s}}$,
as can be seen from Eq.~(\ref{eq:k2k4}). On the other hand, $f_{\mathrm{n}}$
is due to the fluctuations of the individual atomic spins with frequencies
on the order of the exchange coupling $J$.

Further analysis of Figs.~\ref{fig:Spheres}, \ref{fig:Cubes}, \ref{fig:Cubo_octahedron}
reveals further useful information. First, the two frequencies $f_{\mathrm{p}}$
and $f_{\mathrm{c}}$ change with size, shape, and energy parameters.
As the size decreases, the ratio $N_{\mathrm{c}}/\mathcal{N}$ decreases,
and the surface contribution to the overall energy increases~\citep{AdamsEtal_ps23}.
Second, as the surface anisotropy constant $K_{\mathrm{s}}$ increases,
the nutation frequency $f_{\mathrm{n}}$ is more clearly identified
with an increasing amplitude. This effect is more clearly seen in
Figs.~\ref{fig:Spheres} and \ref{fig:Cubo_octahedron} (plots on
the right). However, the value of the frequency itself changes with
the exchange coupling. Likewise, when going from the cube to the more
rounded geometry of the sphere, through the truncated octahedron,
we see that the nutational motion is progressively enhanced and its
amplitude increases.

\subsection{Nutation frequency and amplitude: analytical approach}

From the previous section we conclude that the effective model (EOSP)
perfectly describes the dynamics of the many-spin nanomagnet as far
as the validity conditions are met (\emph{i.e.} small surface anisotropy).
On the other hand, we have seen that, in addition to the nutation
frequency $f_{\mathrm{c}}$, the MSP exhibits another nutation mode
with a much higher frequency $f_{\mathrm{n}}$; the amplitude of this
mode increases with the enhancement of surface effects, either through
an increase of the anisotropy constant or that of the number of surface
spins.

Now, with the macroscopic EOSP model at hand, we can derive analytical
expressions for the frequency and amplitude of the precession and
nutation modes and study their behavior in terms of surface anisotropy
and size. Instead of the Cartesian components $m_{x},m_{y},m_{z}$
of $\boldsymbol{m}$, it is more convenient to parametrize the vector
$\boldsymbol{m}$ using the spherical coordinates $\boldsymbol{q}=\left(\theta,\phi\right)$,
with $\theta$ being the polar angle and $\phi$ the azimuth angle,
\emph{i.e.}, $\boldsymbol{m}\left(\boldsymbol{q}\right)=\left(\sin\theta\cos\phi,\sin\theta\sin\phi,\cos\theta\right)$.
Then, in this coordinate system, the Larmor Eq.~\eqref{eq:NetMagMom_LarmorEqt}
becomes 
\begin{equation}
\frac{d\boldsymbol{q}}{d\tau}=\boldsymbol{b}_{\mathrm{eff}}'\left(\boldsymbol{q}\right),\label{eq:LarmorEqt_SC}
\end{equation}
where we have defined $\boldsymbol{b}_{\mathrm{eff}}'=\left(h_{\theta},h_{\phi}\right)$
as the new effective driving field. More explicitly, the components
of the spherical driving field $\boldsymbol{b}_{\mathrm{eff}}'$ are
given by 
\begin{align}
b_{\mathrm{eff},\theta}'\left(\boldsymbol{q}\right) & =k_{4}\sin\left(4\phi\right)\sin^{3}\theta,\label{eq:SphericalDrivingField}\\
b_{\mathrm{eff},\phi}'\left(\boldsymbol{q}\right) & =2k_{2}\cos\theta-k_{4}\cos\theta[7\cos^{2}\theta-\sin^{2}\theta\cos\left(4\phi\right)-3].\nonumber 
\end{align}
In Eq.~\eqref{eq:SphericalDrivingField}, the appearance of trigonometric
functions with the argument $4\phi$ already hints to the fact that
we should observe a mode with a frequency four times that of the ``fundamental''
mode, namely the precession mode.

To analytically solve the Larmor equation~\eqref{eq:LarmorEqt_SC},
we make the following Ansatz for the angular motion: 
\begin{equation}
\boldsymbol{q}_{p}\left(\tau\right)=\left(\begin{array}{c}
\theta_{p}\left(\tau\right)\\
\phi_{p}\left(\tau\right)
\end{array}\right)=\left(\begin{array}{c}
\theta_{0}\\
\phi_{0}+\nu_{\mathrm{p}}\tau
\end{array}\right),\label{eq:Ansatz}
\end{equation}
where $\omega_{\mathrm{p}}$ is the (circular) precession frequency.
Note that this Ansatz is only valid for $k_{4}=0$. For finite but
small $k_{4}$, we assume that the dynamics is approximately governed
by the equation 
\begin{equation}
\frac{d\boldsymbol{q}}{d\tau}\simeq\boldsymbol{b}_{\mathrm{eff}}'\left[\boldsymbol{q}_{p}\left(\tau\right)\right].\label{eq:LarmorEqt_SC2}
\end{equation}
More explicitly, this yields: 
\begin{align}
\frac{d\theta}{d\tau} & \simeq-k_{4}\sin^{3}\theta_{0}\sin\left(4\nu_{\mathrm{p}}\tau\right),\label{eq:LarmorEqt_SC3}\\
\frac{d\phi}{d\tau} & \simeq-\cos\theta_{0}[2k_{2}+3k_{4}-7k_{4}\cos^{2}\theta_{0}]-k_{4}\cos\theta_{0}\sin^{2}\theta_{0}\cos\left(4\nu_{\mathrm{p}}\tau\right).\nonumber 
\end{align}
From these approximate solutions, we infer the following equations:
\begin{align}
\nu_{\mathrm{p}} & =\cos\theta_{0}\left[2k_{2}+\left(3-7\cos^{2}\theta_{0}\right)k_{4}\right],\label{eq:Freqs-Amps}\\
\nu_{\mathrm{c}} & =4\nu_{\mathrm{p}},\nonumber \\
a_{\phi} & =\frac{1}{4}\,\frac{k_{4}\sin^{2}\theta_{0}}{2k_{2}+\left(3-7\cos^{2}\theta_{0}\right)k_{4}},\,a_{\theta}=a_{\phi}\tan\theta_{0},\nonumber 
\end{align}
where $\nu_{\mathrm{p}}$ and $\nu_{\mathrm{c}}$ are the (dimensionless)
precession and nutation frequencies, and $a_{\theta},a_{\phi}$ the
amplitudes. Note that in situations where $k_{4}$ becomes negative,
the latter should be replaced by $\left|k_{4}\right|$ in the expression
of the amplitude $a_{\phi}$.

\begin{figure*}
\centering \includegraphics[scale=0.21]{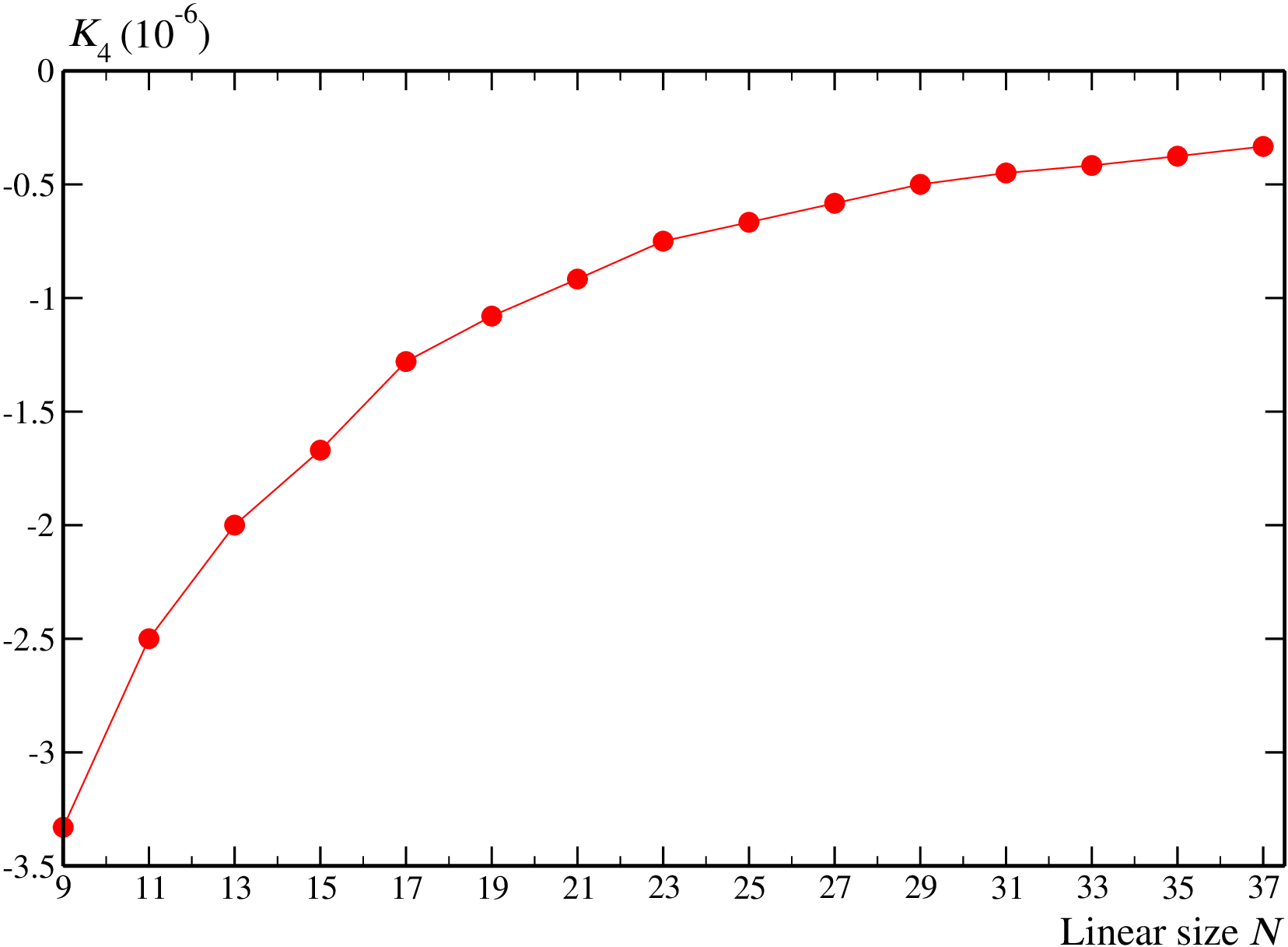}\,\includegraphics[scale=0.205]{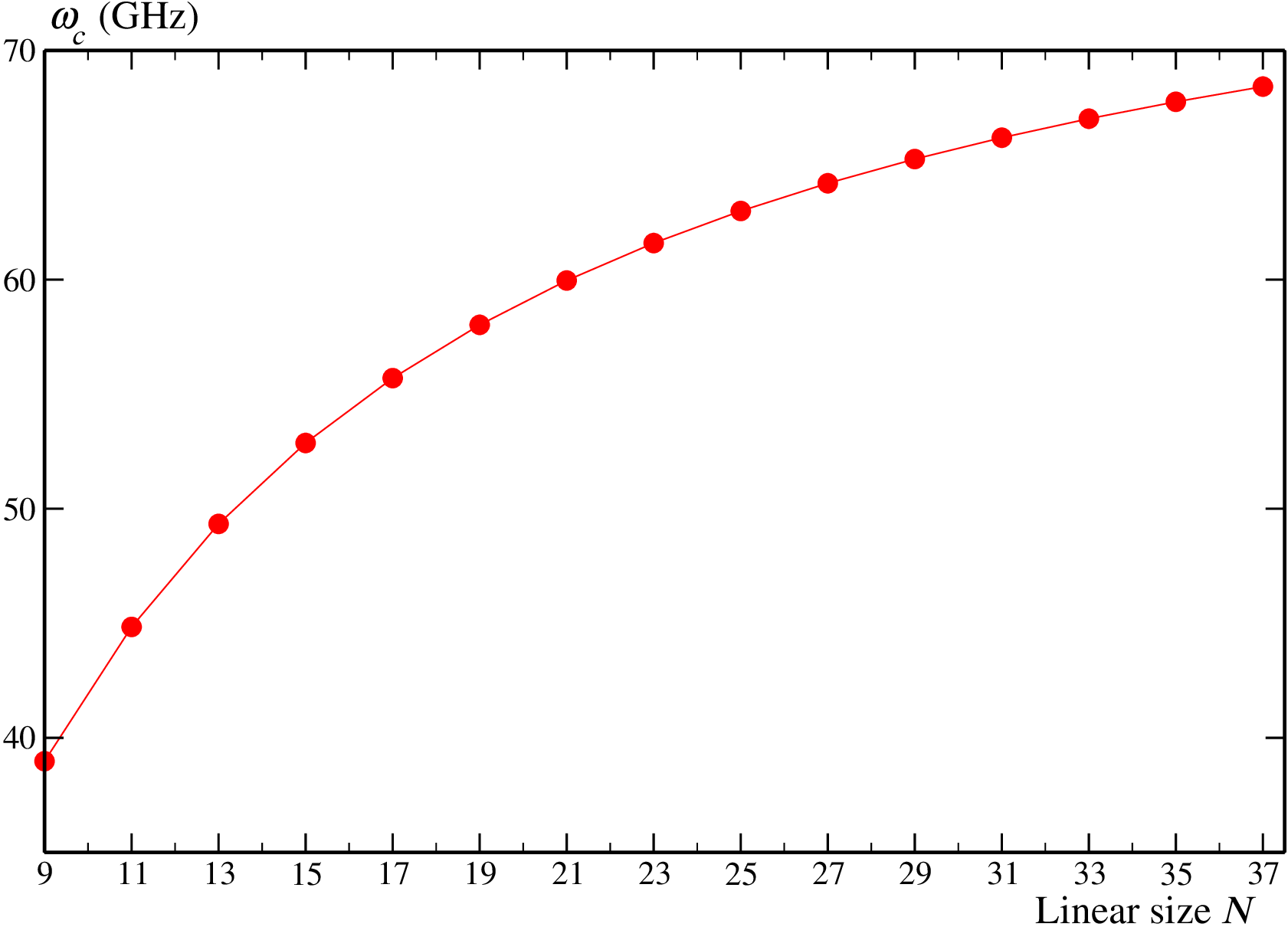}\,\includegraphics[scale=0.21]{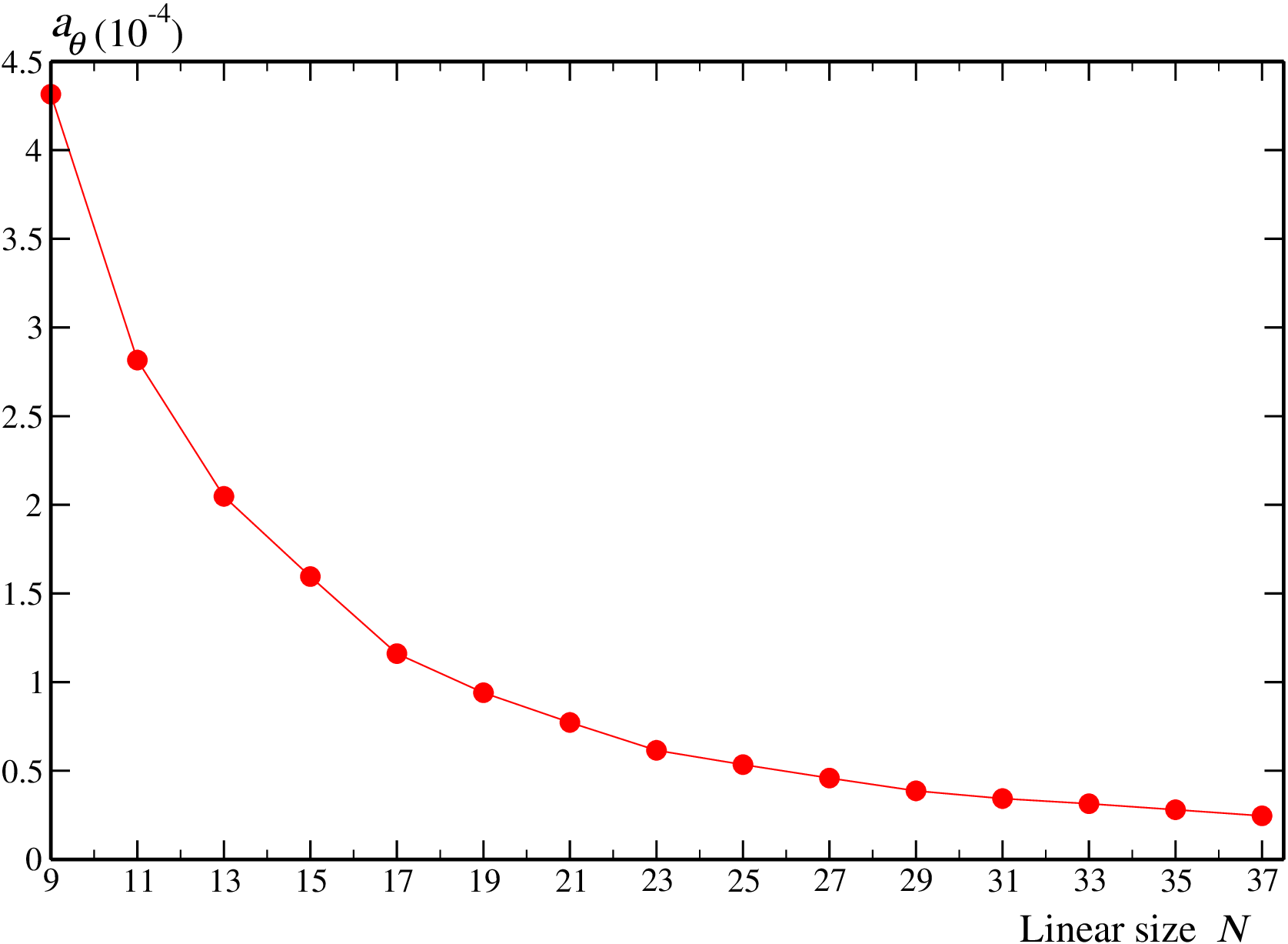}
\caption{\label{fig:Amps-vs-N} EOSP parameter $k_{4}$, nutation frequency
$\omega_{\mathrm{c}}$, and amplitude $a_{\theta}$ as a function
of the linear size $N$ of a bcc iron cube of size $N\times N\times N$
($k_{\mathrm{c}}=0.001$ and $k_{\mathrm{s}}=0.01$). The parameters
were obtained from a fit of the many-spin nanomagnet results to Eqs.~(\ref{eq:Freqs-Amps})
and (\ref{eq:PrecessionNutationApproximation}). The full curves are
guides to the eye.}
\end{figure*}

Integrating Eqs.~\eqref{eq:LarmorEqt_SC3} yields the time trajectories
of the net magnetic moment: 
\begin{align}
\boldsymbol{q}\left(\tau\right) & \simeq\left(\begin{array}{c}
\theta_{0}\\
\phi_{0}+\nu_{\mathrm{p}}\tau
\end{array}\right)-\left(\begin{array}{c}
a_{\theta}\left[1-\cos\left(\nu_{\mathrm{c}}\tau\right)\right]\\
a_{\phi}\sin\left(\nu_{\mathrm{c}}\tau\right)
\end{array}\right).\label{eq:PrecessionNutationApproximation}
\end{align}
The set of Eqs.~(\ref{eq:Freqs-Amps}) and (\ref{eq:PrecessionNutationApproximation})
constitutes the original result of this work. Before discussing the
conclusions we can draw from them, we summarize how they can be used
in practice. Given a (many-spin) nanomagnet of a certain size and
shape and a chosen set of energy parameters $J,K_{\mathrm{c}},K_{\mathrm{s}}$,
so that the spin configuration is not too much disordered, we can
characterize the (modified) precession and the first nutation modes
by their frequencies and amplitudes given by the corresponding effective
macrospin model. If the nanomagnet is a nanocube or a nanosphere with
the underlying simple-cubic lattice, the effective coefficients $k_{2}$
and $k_{4}$ are given by Eqs.~(\ref{eq:k2k4}). In the general case,
one may fit the MSP curves $m_{\alpha}\left(\tau\right)$ to Eq.~\eqref{eq:PrecessionNutationApproximation}
and infer the quantities in Eqs.~\eqref{eq:Freqs-Amps}. In fact,
once the constant $k_{4}$ is obtained, all the other quantities ($\nu_{\mathrm{p}},\nu_{\mathrm{c}},a_{\theta},a_{\phi}$)
can easily be derived. By way of illustration, we have applied the
latter procedure to the much studied iron nanocubes~\citep{tartajetal03jpd,tran06phd,snoecketal08nl,trunovaetal08jap,Respaud_nanocube2_2010JMMM}
with an underlying bcc lattice. The results are shown in Fig.~\ref{fig:Amps-vs-N},
where $k_{4},\omega_{\mathrm{c}}$ and $a_{\theta}$ are plotted against
the (linear) cube size $N$; the fit for $k_{4}$ yields $k_{4}\simeq-12.5k_{\mathrm{s}}^{2}/8\left(1+N\right)^{5/3}$.

From Eq.~\eqref{eq:Freqs-Amps} and the numerical results in Fig.~\ref{fig:Amps-vs-N},
we make the following observations: 
\begin{itemize}
\item The frequencies $\nu_{\mathrm{p}}$ and $\nu_{\mathrm{c}}$ increase
when the size increases but decrease with surface anisotropy, \emph{i.e.}
when $K_{\mathrm{s}}$ ($k_{4}\propto k_{\mathrm{s}}^{2}$) increases,
for a small initial polar angle $\theta_{0}$ (precession angle).
For a relatively large angle $\theta_{0}$ ($\gtrsim49^{\circ}$),
the frequency becomes an increasing function of $K_{\mathrm{s}}$
(plot not shown). 
\item The nutation frequency $\nu_{\mathrm{c}}$ is four times the precession
frequency $\nu_{\mathrm{p}}$, independently of the nanomagnet's size. 
\end{itemize}
Regarding the amplitude, from Eqs.~\eqref{eq:Freqs-Amps} we see
that it increases with the surface anisotropy constant $K_{\mathrm{s}}$,
as already concluded from the results of the MSP simulations in Figs.~\ref{fig:Spheres},
\ref{fig:Cubes}, \ref{fig:Cubo_octahedron}. On the other hand, its
behavior as a function of the size depends on the shape. In Fig. \ref{fig:Amps-vs-N},
we see that it is a decreasing function of $N$.

For a cube with simple-cubic lattice, $N_{\mathrm{c}}/\mathcal{N}=\left(1-2/N\right)^{3}$
and thereby $a_{\phi},a_{\theta}\sim1/N$, which means that when the
size increases, the nutation mode dies out. This has a very important
implication. In general, any bulk or thin magnetic film presents an
anisotropy energy that is an expansion in the components of its net
magnetic moment, similar to the Hamiltonian in Eq.~(\ref{eq:EffPotential}).
However, the dynamics of a bulk magnetic material with such a Hamiltonian
does not necessarily exhibit magnetic nutation. Here, we see that
nanoscaled magnetic systems, such as nanocubes, do develop such a
nutational motion at a frequency that should be observable in resonance
experiments on arrays of such objects. For a sphere, using $k_{2}$
and $k_{4}$ from Eq.~(\ref{eq:k2k4}), Eq.~(\ref{eq:Freqs-Amps})
implies that the amplitudes $a_{\phi},a_{\theta}$ do not depend on
the size, as has also been confirmed by the MSP numerical simulations.
For an arbitrary shape and/or lattice structure, it is not easy to
derive simple formulae for the coefficients $k_{2},k_{4}$. Nevertheless,
as was discussed above, they may be obtained by fitting the dynamics
of the net magnetic moment of the many-spin nanomagnet to Eqs.~(\ref{eq:PrecessionNutationApproximation}).
This is what has been done for the bcc cubic nanomagnets, studied
by experimental groups~\citep{snoecketal08nl,trunovaetal08jap,jiangetal10jac,mehdaouietal10j3m,kronastetal11nl,okellyetal12nanotech,BastardisEtal-jpcm2017},
and the results are shown in Fig.~\ref{fig:Amps-vs-N}.

\section{Conclusion}

We have shown that the magnetization of a nanoscaled magnetic material
may exhibit nutational motion with frequencies in the GHz to THz~range,
the lowest of which is four times the precession frequency. We have
provided analytical expressions for precession and nutation frequencies
and the amplitude of the nutational oscillations as functions of the
size of the nanomagnet and atomistic energy parameters, such as the
spin-spin exchange coupling and the on-site surface and core magneto-crystalline
anisotropies.

This has been possible owing to the correspondence we have firmly
established between the many-spin approach and the effective macrospin
approach, a correspondence that has been validated through numerical
simulations of the dynamics of various nanomagnets (cubes, spheres,
and truncated octahedrons) with different energy parameters and lattice
structures. The results of these simulations have confirmed that the
effective model recovers very well the dynamics of the many-spin nanomagnet
in the low-frequency regime, when surface anisotropy is not too strong.
In this regime, we observe the precession frequency $f_{\mathrm{p}}$
and the lowest nutation frequency $f_{\mathrm{c}}$ that is four times
$f_{\mathrm{p}}$ and whose existence is related to the effective
cubic anisotropy induced by spin inhomogeneities. The many-spin nanomagnet
also exhibits a much higher frequency $f_{\mathrm{n}}$ that can be
related to the spin-spin exchange coupling. Even the precession frequency
$f_{\mathrm{p}}$ turns out to be affected by surface anisotropy,
among other parameters. Within the effective macrospin model, we have
demonstrated that the amplitude of the nutational motion is enhanced
by surface anisotropy, but its behavior with the size depends on the
shape of the nanomagnet.

All in all, in this work, we want to emphasize the fact that surface-induced
magnetization nutation in nanoscaled magnets, such as nanocubes and
nanospheres, already occurs at the lowest frequency that is four times
the precession frequency, which itself is altered by spin misalignments.
This low-frequency nutational oscillation lends itself to a detection
and characterization using standard experimental techniques of magnetic
resonance, such as the network analyzer with varying field and frequency,
at low temperature. An array of well separated platelets or nanocubes
should provide the adequate conditions for such measurements. Inelastic
neutron spectroscopy~\citep{FranzEtAl_jpsj19} might also be able
to resolve the predicted nutational spin dynamics in nanomagnets.
\begin{acknowledgments}
M.P.\ Adams and A.\ Michels thank the National Research Fund of
Luxembourg for financial support (AFR Grant No.~15639149). H. Kachkachi
thanks A.\ Michels and the Department of Physics (Univ. Luxembourg)
for the hospitality extended to him during his stay. 
\end{acknowledgments}

\bibliographystyle{apsrev} %\bibliography{hkbib}

\expandafter\ifx\csname natexlab\endcsname\relax
\global\long\def\natexlab#1{#1}%
\fi \expandafter\ifx\csname bibnamefont\endcsname\relax 
\global\long\def\bibnamefont#1{#1}%
\fi \expandafter\ifx\csname bibfnamefont\endcsname\relax 
\global\long\def\bibfnamefont#1{#1}%
\fi \expandafter\ifx\csname citenamefont\endcsname\relax 
\global\long\def\citenamefont#1{#1}%
\fi \expandafter\ifx\csname url\endcsname\relax 
\global\long\def\url#1{\texttt{#1}}%
\fi \expandafter\ifx\csname urlprefix\endcsname\relax
\global\long\def\urlprefix{URL }%
\fi \providecommand{\bibinfo}[2]{#2} \providecommand{\eprint}[2][]{\url{#2}}

\end{document}